\documentclass{aa}
\usepackage{graphics}

\begin{document}
    

   \title{X-ray Imaging-Spectroscopy of Abell 1835}


   \author{J. R. Peterson \inst{1},
	   F. B. S. Paerels \inst{1},
	   J. S. Kaastra \inst{2},
    	   M. Arnaud \inst{3},
	   T. H. Reiprich \inst{4},
           A. C. Fabian \inst{5},
	   R.~F.~Mushotzky \inst{6},
	   J. G. Jernigan  \inst{7},
           I. Sakelliou \inst{8}}

   \offprints{jrpeters@astro.columbia.edu}
   \authorrunning{Peterson et al.}
   \titlerunning{X-ray Imaging-Spectroscopy of Abell 1835}

   \institute{\inst{1} Columbia Department of Physics and Astrophysics Laboratory, 550 W 120th St., New York, NY 10027, USA\\
	      \inst{2} SRON Laboratory for Space Research Sorbonnelaan 2, 3584
	      CA Utrecht, The Netherlands \\
	\inst{3} CEA/DSM/DAPNIA Saclay, Service d'Astrophysique,  
         L'Orme des Merisiers, B\^at. 709., 91191 Gif-sur-Yvette, France\\
	\inst{4} Max-Planck-Institut f\"ur extraterrestrische Physik,
	P.O. Box 1312, 85741 Garching, Germany \\
	\inst{5} Institute of Astronomy, Madingley Road, Cambridge CB3 0HA, UK \\
	\inst{6} NASA/GFSC, Code 662, Greenbelt, MD, 20771, USA \\
	\inst{7} Space Sciences Laboratory, University of California, Berkeley, CA 94720, USA \\
	\inst{8} Mullard Space Science Laboratory, UCL, Holmbury St. Mary,
	Dorking, Surrey RH5 6NT, UK \\}

   \date{Submitted September 29, 2000 / Accepted October 25, 2000 }


	\abstract{	We present detailed spatially-resolved spectroscopy results of the observation of
      Abell 1835 using the European Photon Imaging Cameras (EPIC) and the
      Reflection Grating Spectrometers (RGS) on the XMM-Newton observatory.
      Abell 1835 is a luminous ($10^{46}$ ergs $\mbox{s}^{-1}$), medium redshift ($z=0.2523$), X-ray emitting cluster
      of galaxies.  The observations support the interpretation that
      large amounts of cool gas are present in a multi-phase medium surrounded
      by a hot ($kT_{e}$=8.2 keV) outer envelope.  We detect O VIII Ly$\alpha$
      and two Fe XXIV complexes in the RGS spectrum.  The emission measure of the cool gas
      below $kT_{e}$=2.7 keV
	 is much lower than expected from standard cooling-flow models,
      suggesting either a more
      complicated cooling process than simple isobaric radiative
      cooling or differential cold absorption of the cooler gas.
}

   \maketitle

\section{Introduction}

\begin{figure} 
  \resizebox{8.5cm}{!}{\rotatebox{0}{\includegraphics{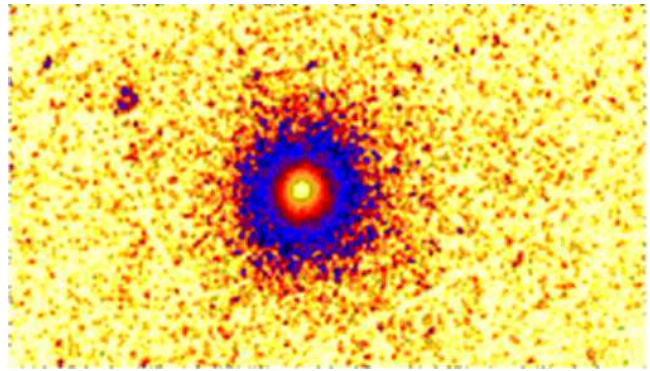}}}
      \caption{Smoothed logarithmic intensity EPIC-MOS image of Abell 1835.
      The image is approximately 23 by 13
      arcminutes.}
  \label{fig:f1}
\end{figure}

Previous soft X-ray imaging and medium spectral resolution studies have
implied the presence of cool gas near the centers of clusters (See
e.g. Sarazin \cite{Sarazin88}, Fabian \cite{Fabian94a}).  For the Centaurus,
Perseus, Virgo, and Coma clusters there is
direct spectroscopic evidence for cool gas from the {\it Einstein} Focal Plane Crystal
Spectrometer (Canizares et al. \cite{Can79}, \cite{Can82}, Mushotzky \& Szymkowiak \cite{Mushot88}).  Based on
simple, robust estimates for the electron densities and temperatures,
it is estimated that in the large majority of clusters, gas in the
central volume should cool radiatively to very low temperatures on a
timescale short compared to the characteristic age of the cluster.
Due to the consequent lack of central pressure support, gas has to
continuously flow inward into the cluster potential. 

This picture of cluster cooling flows is incomplete in that the
fate of the cooling gas is unknown. Material that
has cooled to below X-ray emitting temperatures generally fails to be
detected by emission at any other wavelengths, although more recently
there are indications that cold gas may be distributed throughout the
cooling flow. This gas betrays its presence through X-ray absorption
of the cooling flow emission (but see below for evidence that such
absorption is not confirmed at higher spectral resolution).  A simple
single phase model for the cooling gas based on the 
conservation laws for the global flow predicts a sharply peaked 
density distribution that is not consistent with the observed
distributions, and this fact has given rise to the idea of a diffuse,
multi-phase cooling flow (Johnstone et al.\cite{Johnstone92}).  

Abell 1835 is a prototypical strong cooling-flow cluster with an inferred mass
deposition rate of $1750 \pm 550$ (Allen et al. \cite{Allen96}).  The central
cooling time (the time it would take the gas to radiate away all of its kinetic energy) is less than 1 Gyr, whereas the cluster could remain relatively
undisturbed by mergers for $\sim$ 10
Gyr.  So roughly the inner 350 kpc of the cluster should have had sufficient
time to cool below X-ray temperatures.
Allen et al. (\cite{Allen96}) found Abell 1835 to be one of the strongest cooling-flows
from the deprojection of ROSAT HRI images.  The ASCA spectrum was consistent
with a $kT_e=8.5~\mbox{keV}$ collisionally ionized plasma, but cooling-flow spectral
models were unconstrained (Allen et al. \cite{Allen96}).
Allen and Fabian (\cite{Allen98}) also carried out a joint ROSAT-ASCA study
and found evidence for a massive cooling flow, as well as a relatively high metallicity.  Until this observation,
cooling-flow properties have largely been inferred by relatively short cooling
times.  Here we present spectroscopic results on the properties of the cool gas.

As part of the XMM-Newton (Jansen et
al \cite{Jansen00}) performance verification program, Abell 1835 was observed for 60 ks.
Data were obtained simultaneously with the two Reflection Grating Spectrometers (RGS) and
the three
European Photon Imaging Cameras (EPIC).  The RGS is capable of achieving high
spectral resolution for the entire soft wavelength band (5 to 38 $\mbox{\mbox{\AA}}$)
for moderately extended X-ray sources.  The spectral resolution for sources of
characteristic angular size, $\theta$, is roughly, $\Delta \lambda \approx 0.1
~\mbox{\AA}~(\theta / 1^{\prime}) $. 
The high throughput of the XMM
mirrors allows the EPIC-pn and EPIC-MOS detectors to perform detailed spectral-spatial
studies of clusters since the spatial extent of most clusters is resolved.
Both sets of instruments can be used together to produce a spectral-spatial model of the X-ray emission.

Throughout this paper we assume $H_0= 70~\mbox{km}~\mbox{s}^{-1}~\mbox{Mpc}^{-1}$, $\Omega_m =
0.3$, and $\Omega_{\Lambda}=0.7$.  This implies an angular distance of 1280
Mpc and a luminosity distance of 1600 Mpc for Abell 1835.  One arcsecond
corresponds to 6.2 kpc. Errors are quoted at the $90\%$ confidence level and
systematic errors are added in quadrature whenever they can be estimated.

\section{Overall Temperature Structure}

In order to study the cooling flow properties of Abell 1835 in detail, we
first use the spatially-resolved EPIC-MOS and EPIC-pn spectra to characterize the
temperature and density structure.  The  EPIC instruments
and their calibration are described in Turner et al. (\cite{Turner00}) and
Str\"uder et al. (\cite{Strueder00}).
An image of Abell 1835 is shown in Fig. 1.  

We performed spectral fits with EPIC-MOS and EPIC-pn on two concentric annuli in Abell 1835.  EPIC-pn fits are shown in Figs. 2 and 3.  Background spectra were used from Lockman Hole observations
(Revolution 70 and 73, 100ks observation) for the MOS fits or off-source
regions and the PN fits.  The spectra were corrected
for energy-dependent telescope vignetting. (See also Arnaud \cite{Arnaud00})
The spectral and spatial responses do not vary strongly over the chosen
annuli.   Periods with high background rates were removed.
We fit the spectra to single temperature models with the XSPEC v11 package
(Arnaud \cite{Arnaud96}) using the MEKAL (Mewe, Kaastra, \& Liedahl \cite{Mewe95}) spectral model.
Elemental abundance ratios
 were held fixed to their solar values when they were not well-constrained.

Within a 0.5 arcminute radius circle, Abell 1835 has a temperature of $\sim6 ~\mbox{keV}$ and an iron abundance of $0.35$ relative to solar.  The
outer region,  $0.5 < r < 2$ arcminutes, has a temperature of $\sim8 ~ \mbox{keV}$
and an iron abundance of $0.28$ of solar values.  Column densities were consistent
with the galactic value in the MOS fits and slightly
lower than galactic in the PN fits.  There is no evidence for a strong
abundance gradient.  The results of the fits are summarized in
Tables 1 and 2. 

The spectral fits are consistent with the interpretation that there is a large
volume in the center of Abell 1835 with cooler gas at least as cool as 6 keV.
We use
the single temperature fits to the outer annulus to provide an upper
temperature limit for the RGS analysis.

   \begin{table}
      \caption{Basic Temperature Structure: EPIC-MOS Spectral Fits}
         \label{EPIC}
      \[
         \begin{array}{p{0.5\linewidth}ll}
            \hline
            \noalign{\smallskip}
Parameter & \theta < 0.5^{\prime} & 0.5^{\prime} < \theta < 2^{\prime} \\ 
            \noalign{\smallskip}
            \hline
            \noalign{\smallskip}
kT (keV)  & 6.0 \pm 0.3      & 8.2\pm0.4   \\ 
$A_{\mbox{Fe}}$ & 0.35 \pm0.05  & 0.28\pm 0.05  \\ 
$N_{\rm{H}} (10^{20}\mbox{cm}^{-2})$ & 2.6\pm0.4     &2.1\pm0.4   \\ 
$\chi^{2}/\mbox{dof}$ & 1.08   & 0.97   \\ 
            \noalign{\smallskip}
            \hline
         \end{array}
      \]
   \end{table}

   \begin{table}
      \caption{Basic Temperature Structure: EPIC-pn Spectral Fits}
         \label{PN}
      \[
         \begin{array}{p{0.5\linewidth}ll}
            \hline
            \noalign{\smallskip}
Parameter & \theta < 0.5^{\prime} & 0.5^{\prime} < \theta < 2^{\prime} \\ 
            \noalign{\smallskip}
            \hline
            \noalign{\smallskip}
kT (keV)  & 5.7 \pm 0.2      & 8.4\pm0.4   \\ 
$A_{\mbox{Fe}}$ & 0.34 \pm0.05  & 0.28\pm 0.06  \\ 
$N_{\rm{H}} (10^{20}\mbox{cm}^{-2})$ & 1.8\pm0.3    &1.2\pm0.3   \\ 
$\chi^{2}/\mbox{dof}$ & 1.11   & 0.98   \\ 
            \noalign{\smallskip}
            \hline
         \end{array}
      \]
   \end{table}

\begin{figure} 
  \resizebox{7cm}{!}{\rotatebox{-90}{\includegraphics{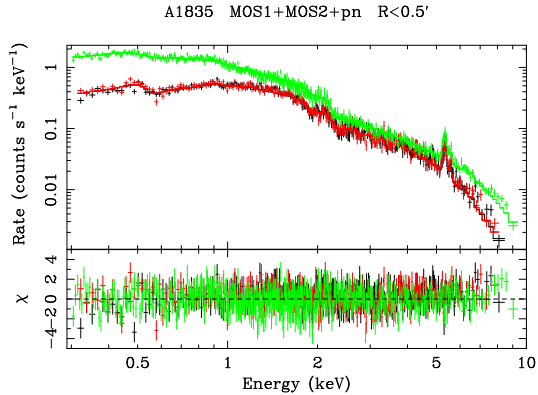}}}
  \caption{EPIC-pn (green), EPIC-MOS 1 (black) and EPIC-MOS 2 (red) spectral fit and residuals of the inner region of Abell 1835.}
  \label{fig:f4}
\end{figure}  

\begin{figure} 
  \resizebox{7cm}{!}{\rotatebox{-90}{\includegraphics{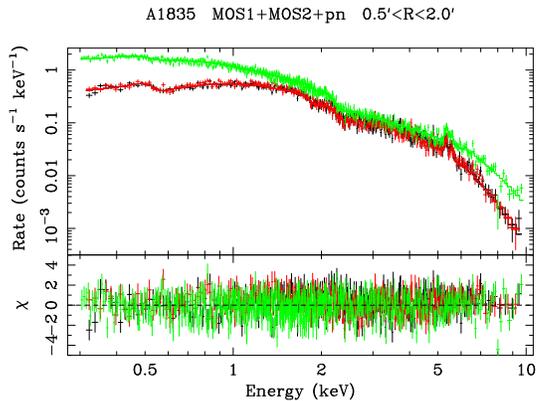}}}
  \caption{EPIC-pn (green), EPIC-MOS 1 (black) and EPIC-MOS 2 (red) spectral fit and residuals of the outer region of Abell 1835.}
  \label{fig:f5}
\end{figure}

\section{Cooling-Flow Spectroscopy}

By studying the RGS spectrum, we can measure both the relative emission
measure of cold gas and study the spectral properties of the cooling flow in
detail.

The properties of the RGS and its calibration are described by den Herder et
al. (\cite{JW00}).  The RGS instrument effectively
works as a long-slit spectrometer, providing spectral information in one
direction (somewhat degraded by the spatial extent of the source) while
providing 15 arcsecond imaging information in the cross-dispersion direction.  
For each photon, the focal plane detectors record the position along the
dispersion direction (which gives the dispersion angle), the position in the
cross-dispersion direction, and the CCD pulse-height.  An energy dependent
joint dispersion coordinate/pulse-height filter separates the spectral orders.
In this analysis, we jointly use first and second orders.   For all spectra
presented here, we assign wavelengths based on the nominal position of
the center of the X-ray emission ($\alpha=14^h 1^m 1^s$, $\delta=+2^{\circ}
52^{\prime} 43^{\prime}$).

The spectrum for both RGS instruments is shown in Fig. 4.  
The spectrum has been corrected for redshift, divided by the effective area
and exposure, and a background model is subtracted for display purposes only.
These corrections have been applied based on a Monte Carlo simulation described below.
Time intervals with background rates more than 4
times higher than the quiescent rate were removed from the analysis to
maximize the signal to noise in the RGS instrument.   For the data selection
cuts used here, the background comprises $\sim 30\%$ of the counts in the
extracted spectrum.  Approximately $\sim 70\%$ of the 60 ks observation was
used in this analysis.

We detect two Fe L complexes dominated by Fe XXIV lines ($10.6~\mbox{\AA}$, $11.4~\mbox{\AA}$) and O
VIII Ly $\alpha$ ($18.97\mbox{\AA}$) above the bremsstrahlung continuum.  
The lines are broadened by a sharply peaked intensity distribution of $\sim$ 1
arcminute FWHM.  There are some non-statistical fluctuations, but all of these
are much narrower than true emission lines from the cooling-flow would appear,
and none of them are coincident with known strong astrophysical emission lines.
The presence of the Fe XXIV ion suggests cool gas with a
temperature between 1 and 3 \mbox{keV}, whereas O VIII could have contributions
from both cool ($\sim 1~\mbox{keV}$) and hot gas ($\sim 10~
\mbox{keV}$).  The absence of other Fe L shell ions produces strong constraints on cooling-flow models as explained below.

To fully address spectral-spatial coupling in the RGS, we compare the data
directly to models by folding the spectral-spatial models through an
instrument response Monte Carlo.  The Monte Carlo takes wavelengths and sky
positions of photons, and predicts cross-dispersion coordinates, dispersion
coordinates, and CCD pulse-heights.  The Monte Carlo includes characterizations of
all known physical properties of the RGS instrument at the time of the writing of this
paper, with the exception of a not yet calibrated O K instrumental edge
of approximately 20$\%$ at 23.5 $\mbox{\AA}$, which has moved to 18.7
$\mbox{\AA}$ in our correction of the redshift of Abell 1835 in Fig. 4.  The analysis is then unaffected by the details of data selection cuts
provided the same cuts are applied to the data and the Monte Carlo simulated
photons.  The Monte Carlo properly models the off-axis response of the RGS.  The cluster emission is distributed across the entire CCD array so
there is no source-free region to estimate the background contribution.
Instead, we use a model based on the Lockman Hole (Revolution 70 and 73,
100ks observation)
which characterizes the effects of low energy protons, the calibration
sources, and the CCD detector noise.  The background is smooth and is
relatively flat across the wavelength band.

In order to characterize spatial structure for use in the RGS Monte Carlo, we
use a previous ROSAT observations for the density parameterization, which is
consistent with the observed EPIC spatial distribution.
We use a $\beta$ model (King
\cite{King66}, Cavaliere \& Fusco-Fermiano \cite{Cavaliere76}) to describe the density structure. The model has a core radius of 18 arcseconds, $\beta$ of 0.72.  This
characterization reproduces the RGS cross-dispersion event distribution.   The exact spatial
distribution is unimportant since it only affects the shape of the spectral line
profiles and only a rough spatial characterization is needed.
For the spectral models we use
the coronal plasma MEKAL model of Mewe, Kaastra, and Liedahl (\cite{Mewe95}), the cold absorption
model of Morrison and McCammon (\cite{Morrison83}), and a spectral-spatial
renormalization of the MKCFLOW model for the cooling-flow emission measure
included in the XSPEC v11.0 package (Arnaud \cite{Arnaud96}).

   \begin{table}
      \caption{Cooling Flow Properties: RGS Monte Carlo$^{\rm{a}}$}
         \label{RGS}
      \[
         \begin{array}{llll}
            \hline
            \noalign{\smallskip}
$\mbox{Parameter}$ & $\mbox{Isothermal}$ & $\mbox{Cooling-}$ &
      \mbox{Cooling-}  \\
& & \mbox{Flow } 1 & \mbox{Flow } 2 \\
            \noalign{\smallskip}
            \hline
            \noalign{\smallskip}
kT_{\mbox{ambient}} (\mbox{keV})  & 8.2 & 8.2 & 8.2 \\ 
A_{\mbox{Fe}} & 0.35  & 0.35 & 0.35  \\ 
N_{\rm{H}} (10^{20}\mbox{cm}^{-2}) & 2.32 & 2.32 & 2.32  \\
\theta_C (\mbox{arcseconds}) & 18 & 18 & 18 \\
\beta & 0.72 & 0.72 & 0.72 \\
r_{\mbox{cool}} (\mbox{kpc}) & .. & 180 & 180 \\ 
A_{\mbox{O}} & 0.31\pm0.07 & 0.19\pm0.04 & 0.28\pm0.07 \\
\dot{M} (M_{\sun}~\mbox{yr}^{-1}) & .. & 1700 & 2300  \\
kT_{\mbox{min}} (\mbox{keV}) & .. & .. & 2.7 \pm 0.2 \\
N_{\rm{H}}^0 (10^{20}\mbox{cm}^{-2}) & .. & 0.^{+10.0}_{-0.0} & 0.^{+6.0}_{-0.0} \\ \hline
\mbox{Reduced~} \chi^2~^{\rm{b}} & 1.08 & 2.34 & 0.92 \\ 
\mbox{Reduced~} \chi^2~^{\rm{c}} & 1.59 & 3.46 & 1.11 \\ 

            \noalign{\smallskip}
            \hline
         \end{array}
      \]
\begin{list}{}{}
\item[$^{\rm a}$] Parameters with no errors were held fixed
\item[$^{\rm b}$] For the entire wavelength band (288 degrees of freedom)
\item[$^{\rm c}$] For the 10 to 18 \AA~region (100 degrees of freedom)
\end{list}
   \end{table}

We select the spectral model for the Monte Carlo simulation as a simple isothermal model folded through the instrument
response based on the 8.2 $\mbox{keV}$ measured temperature from the EPIC-MOS
fit for the ambient gas.  We also add a background component.  We set all
elemental abundances other than that of oxygen to ${1}/{3}$ of solar values
and we set the iron abundance to 0.35 of solar values.  The spatial distribution is given by the above mentioned beta model.  We see that the
agreement with the measured continuum is quite good (see Fig. 4).  The oxygen line flux is
well-matched by adjusting the abundance to $0.31\pm0.07$ of the solar value.  The Fe XXIV complexes are 
above the expectation for the continuum and are directly attributable to the cooling-flow.

Next, we attempt a multi-phase cooling flow model where the ambient gas is fixed
at a temperature of 8.2 keV and a fraction of the cool gas is expected to
exist at all radii at several temperatures. 
Following the model of Johnstone et al. (\cite{Johnstone92}),
we model the cool component as having the same spatial distribution as the
ambient component but only existing within the cooling radius.  We have
insufficient counts in the spectral lines to determine the spatial
distribution of the cool gas relative to the hot ambient gas.
The emissivity of the cooling-flow gas is calculated from a conventional
isobaric cooling-flow prescription, in which each temperature interval is
weighted with the time the gas spends cooling through that interval.  The
resulting differential emission measure is then proportional to the inverse of
the total radiative cooling function.
  We also allow for additional cold absorption applying it only
to the cooling flow component at the redshift of the source, where the flux is
reduced by $(1-e^{-\tau})/\tau$ to account for self-absorption.  This would
apply if large deposits of cold byproducts of the cooling flow exist uniformly
in the
cooling-flow region.  We fix the foreground column density to the galactic value and
apply it to both spectral-spatial components.  The relative
normalization between the two components is determined by the mass deposition
rate ($\dot{M}$) and the overall normalization is determined by central
electron number density.

To match the observed equivalent widths of the Fe XXIV lines (10.66 $\mbox{\AA}$, 10.62
$\mbox{\AA}$, 11.03 $\mbox{\AA}$, 11.18 $\mbox{\AA}$, 11.43 $\mbox{\AA}$) ,
we require a mass loss rate of 1700 $M_{\odot}~\mbox{yr}^{-1}$, but the model
predicts significant line emission from several other iron
ions that is not observed.  This is seen quite easily from the blue model in the insert in Fig
4.  In particular, Fe XVII (15.01 $\mbox{\AA}$, 17.05 $\mbox{\AA}$,
17.10 $\mbox{\AA}$, 16.78 $\mbox{\AA}$) lines are most notably
absent.  
Other ions such as O VII (21.6, 21.8, 22.1 $\mbox{\AA}$) are not
observed, but are not expected to be significant since the gas should cool
very rapidly at lower temperatures.  

Previous ASCA studies found an emission line deficit at low energies (Fabian
\cite{Fabian94}), but this was modeled as uniformly distributed additional absorption of
$\sim10^{21} \mbox{cm}^{-2}$ (Allen et al. \cite{Allen00}).
Here we find no evidence for additional absorption uniformly distributed  ($N_{\mbox{H}} < 6\times 10^{20}
\mbox{cm}^{-2}$).  A larger uniform column density would have produced detectable
absorption of the continuum seen through the cooling flow.  Any large
absorption also reduces the flux of the Fe XXIV lines.  We do not see this
absorption so we model the spectrum by cutting off the temperature distribution.
The EPIC fits to cooling flow models required either additional cold absorption of $\sim10^{21}~\mbox{cm}^{-2}$ or a cut-off in
the temperature emission measure distribution at $\sim2~\mbox{keV}$.

To characterize the spectrum empirically, we use the same cooling-flow model
but put a lower temperature cut-off on the emission
measure distribution of $2.7~\mbox{keV}$ as shown by the green model in Fig. 4.  This cut-off is strongly
constrained by the lack of Fe XXIII emission.  The Fe XXIV 10.6
$\mbox{\AA}$ complex is not well-matched because the true emission measure
distribution probably falls more rapidly than implied by the isobaric cooling
model.  Fe XXIII contributes partly to the 11.4 $\mbox{\AA}$ complex (Savin \cite{Savin96}).  

If an additional cooling-flow component is added
with no cutoff on the emission measure distribution, then its mass loss rate
is less than 200 $M_{\odot}~\mbox{yr}^{-1}$ ($90\%$ confidence+systematic).  The strongest constraint comes from the Fe
XXIII, Fe XII, Ne X complex at $12.2 \mbox{\AA}$.  A smaller mass flow rate may be more consistent
with the detection of molecular gas from CO emission, an indicator of star
formation, of $1.4 \times 10^{11}~\mbox{M}_{\sun}$. (A. Edge, private
communication) Similarly, Allen ({\cite{Allen95}) and Crawford et al. (\cite{Crawford95}) find optical
spectra consistent with 100 $M_{\odot}~\mbox{yr}^{-1}$.
 Another interpretation
of the data would be that a very small fraction ($5\%$) of the cooling region
is actually cooling and gas at intermediate temperatures coexists, which
accounts the Fe XXIV lines.  

\begin{figure*} 
  \resizebox{18cm}{9cm}{\rotatebox{90}{\includegraphics{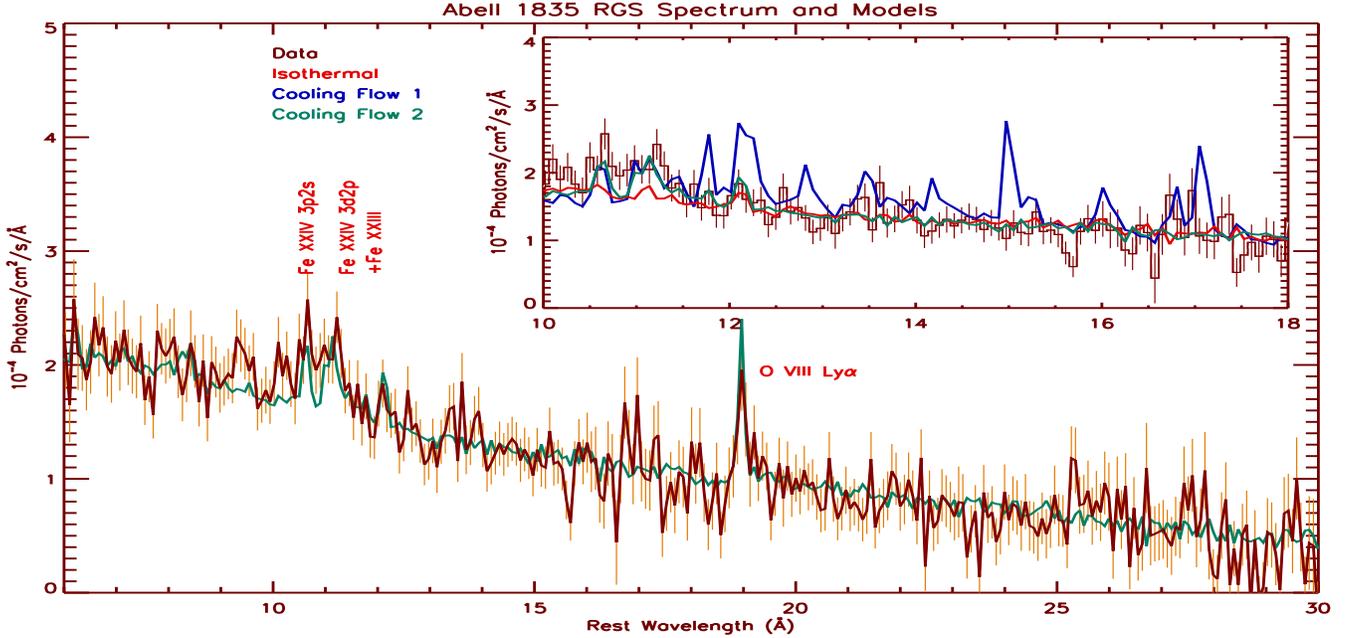}}}
      \caption{RGS spectrum of Abell 1835 and three models folded through
      the instrument Monte Carlo.  The same corrections and data selection
      cuts have been applied to the data and the simulated photons.  The red
      model is an isothermal 8.2 $\mbox{keV}$ model.  The blue model has an
      hot ambient 8.2 $\mbox{keV}$ component and an isobaric cooling flow
      component.  The green model is the same as the blue model but does not
      have emission below 2.7 $\mbox{keV}$.  The details of each model are
      described in the text.  The spectrum is corrected for redshift,
      exposure, and effective area.}
  \label{fig:f7}
\end{figure*}

\section{Discussion}

The emission measure of the cooling-flow spectral model has to be artificially
cut off at about 2.7 $\mbox{keV}$.  This would be possible if the cooling
time was reasonably long compared with the expected age of the cluster.  There
are three possibilities for this lack of emission.  1. It
possible that there is an additional transport mechanism that has
tempered the cooling and the becomes a {\it frustrated } cooling-flow quickly
reheating the cool gas back to the the hot phase.
2.  The gas has rapidly cooled below 2.7 $\mbox{keV}$ and we
therefore do not see it.  3. The cool gas may exist, but its emission is absorbed by cold material at the center of the flow.
It is also important to note that the other XMM-Newton observations of clusters show very
similar results suggesting that this is a generic problem in cooling-flow
clusters (see Kaastra et al. \cite{Kaastra00}, Tamura et al. \cite{Tamura00}).
Following, we discuss some possibilites.

{\bf Underestimate of the Cooling Function:}  The presented results depend
somewhat on the cooling function assumed below $\sim 1 ~\mbox{keV}$, but the
function would have to be underestimated by a factor of at least 10 to explain
the results.  However, for a gas at $kT_e$=1 keV, we detect
primary radiative cooling directly in the X-ray emission lines in our band.
From deep observations of coronal sources (Brinkman et al. \cite{Brinkman00}), we
know that the emission spectrum is accurately accounted for by the current
spectral models, and hence the cooling function does not have a large error.
Our results depend, however, on the metals IGM being approximately
homogeneously distributed.

{\bf Resonant Scattering + Cold Absorption:} This would be
expected to be important in spectral lines with high oscillator strength where
the optical depth, $\tau \sim$ 10 (e.g. 15 \AA~Fe XVII could be $\tau \sim$ 100).
Resonance line photons could then scatter several times within the medium,
which would appear to enhance the
probability of photoelectric absorption.   However, because resonance
scattering of a single transition also has the effect of scattering the photon
in frequency, the photoelectric absorption probability is not enhanced by a
large factor.
Empirically, lines with smaller oscillator strength, which are not subject to
resonance scattering at all, are not observed (e.g. 17.1
$\mbox{\AA}$).   The $10.6~\mbox{\AA}$ complex from Fe XXIV contains two
medium oscillator strength lines which are still observed.

{\bf Differential Cold Absorption:}  A column of $5\times 10^{21}
\mbox{cm}^{-2}$ being selectively applied to the gas below 3 keV would explain
the observed spectrum.  The limit of $6\times 10^{20}~\mbox{cm}^{-2}$ on the column density
being applied to the hot gas, however, would make this a very specific partial
covering model.  This could be reasonable, since the coolest gas occupies the
smallest volume and is likely to be near the coldest material.  The data,
however, are inconsistent with the entire cooling volume being occupied by cold
material.  Further studies may help to constrain the mass and location of the
cold material as well as the volume fraction of the cooling-flow.

{\bf Non-Equilibrium Ionization:}  This effect would cause most ions to
radiate at a lower temperature than is expected from collisional equilibrium
models. (Edgar and Chevalier \cite{Edgar86}).  The recombination time scale for Fe XVII at $10^7~\mbox{K}$ is
$(\alpha_r n_e)^{-1} \approx (10^{-12}~\mbox{cm}^{3}~\mbox{s}^{-1} \times 10^{-2}~\mbox{cm}^{-3})^{-1} \approx 2 \times 10^{6}~\mbox{yr}$ (Arnaud \cite{Arnaud92}).  The cooling time, however, is $\approx 10^{9}~\mbox{yr}$, so this is unlikely to be important.

{\bf Evolutionary Effects}:  The cooling time argument assumes that
evolutionary effects are unimportant.  A recent cluster merger, for example,
would cause an overestimate of the cooling radius.  This seems unlikely since
we would have to be fortunate enough to see the cooling-flow just beginning.
Another possibility is that there is secular mass injection throughout the
lifetime of the cluster.  This also seems unlikely since the mass of the IGM
is larger than the mass of the galaxies themselves.

{\bf Non-Isobaric Cooling-Flow}:  The cooling time estimate is based on the
assumption that the gravitational potential can be neglected.    The
cooling-flow will be non-isobaric if it is flowing into a region of a greater
gravitational potential.  This could apply if the gravitational potential does
significant work on the cool gas (see e.g. Nulsen et al. \cite{Nulsen82}, Wise \cite{Wise93}).  This would cause
an overestimate of the mass deposition rate and adds energy to the gas if the
cooling-flow actually flows.

{\bf Electron Thermal Conduction}:  
Theoretical calculations have disagreed about how much conduction is supressed
by twisted magnetic fields (see
e.g. Tao (\cite{Tao95}) and Chandran and Cowley (\cite{Chandran98})) and
observationally it appears to be suppressed (Ettori and Fabian \cite{Ettori00}).
Electron thermal conduction would have to be suppressed enough to allow for
the observed cooling and temperature gradients, but would have to
play a role in limiting cooling.   Note that cooling works best to balance
cooling at higher temperatures.

{\bf Magnetic Reconnection}:  If large scale magnetic fields were present in
clusters, they might have the effect of isolating cooling flow regions as well
as reconnecting cooling-flow regions with the ambient gas (Soker and Sarazin
\cite{Soker90}, Norman and Meiksin \cite{Norman96}).  This would have the
effect of recycling the cool material and delaying the cooling time.  

{\bf Turbulent Mixing}:  Begelman and Fabian (\cite{Begelman90}) suggested that
turbulent mixing could result in an intermediate temperature layer between the
hot gas and the cold condensed cooling-flow byproducts.  If this process were
highly efficient the cold gas could get recycled into an intermediate phase
and the hot gas could be cooled rapidly by turbulent mixing within the cold
layer. 

{\bf Dust Sputtering: }  Abell 1835 is an IRAS source (Allen \cite{Allen00}).
This could imply that the dust is mixing with the cooling-flow and rapidly
cooling the gas.  

The ideas above are only a sample of the possible mechanisms for the absence
of the 1 keV cooling-flow gas and more than one could be important.  The
observational evidence places strong constraints on cooling flow models and
proper simulations of these effects are beyond the scope of this paper.  With
future XMM-Newton observations using spatially-resolved spectroscopy and or further
limits or detections of Fe XVII of many clusters, we expect to be able to distinguish between the alternatives.  

\section{Conclusion}
	We describe the  spectral-spatial results of the observation of
      Abell 1835 using the European Photon Imaging Cameras and the
      Reflection Grating Spectrometers on the XMM-Newton observatory.
      The observations are consistent with 
      large amounts of cool gas surrounded by 
      a $\mbox{kT}_e$=8.2 $\mbox{keV}$ component outer envelope.  We find no
      evidence for excess uniform cold absorption above the
      galactic value of $2.32\times10^{20} \mbox{cm}^{-2}$.  The emission
      measure of cool gas
      below $\mbox{kT}_e$=2.7 $\mbox{keV}$ is much lower than expected in
      simple isobaric cooling flow
      models, and it indicates either differential cold
      absorption applied to the cool gas or that a more complicated description of the cooling process is required.

\begin{acknowledgements}

This work is based on observations obtained with XMM-Newton,
an ESA science mission with instruments and contributions directly funded by ESA Member States and the USA (NASA).

\end{acknowledgements}

\end{document}